\documentstyle[prl,aps,epsf,psfig]{revtex}
\begin{document}
\twocolumn[\hsize\textwidth\columnwidth\hsize
           \csname @twocolumnfalse\endcsname
\title{Virial corrections to simulations of heavy ion reactions}
\author{Klaus Morawetz}
\address{Fachbereich Physik, University Rostock,
D-18055 Rostock, Germany}
\author{V\'aclav \v Spi\v cka, Pavel Lipavsk\'y}
\address{Institute of Physics, Academy of Sciences, Cukrovarnick\'a 10,
16200 Praha 6, Czech Republic}
\author{Gerd Kortemeyer}
\address{NSCL, Michigan State University,
East Lansing, MI 48824-1321, USA}
\author{Christiane Kuhrts}
\address{Fachbereich Physik, University Rostock,
D-18055 Rostock, Germany}
\author{Regina Nebauer}
\address{SUBATECH,  
Laboratoire de Physique Subatomique,  
4, Rue Alfred Kastler La Chantrerie,  
44070 Nantes Cedex 03,  
France}
\maketitle
\begin{abstract}
Within QMD simulations we demonstrate the effect of virial corrections
on heavy ion reactions. Unlike in standard codes, the binary collisions
are treated as non-local so that the contribution of the collision flux
to the reaction dynamics is covered. A comparison with standard QMD
simulations shows that the virial corrections lead to a broader proton
distribution bringing theoretical spectra closer towards experimental
values. Complementary BUU simulations reveal that the non-locality
enhances the collision rate in the early stage of the reaction. It
suggests that the broader distribution appears due to an enhanced
pre-equilibrium emission of particles.
\end{abstract}
\vskip2pc]

The Boltzmann equation including the Pauli blocking (the BUU equation
\cite{BD88}) and the closely related method of quantum molecular
dynamics \cite{SG86,A91} (QMD) are extensively used to interpret
experimental data from heavy ion reactions. Due to their quasi-classical
character, they offer a transparent picture of the internal dynamics of
reactions and allow one to link observed particle spectra with
individual stages of reactions.

An ambition to cover the heavy ion reactions within experimental errors
has been recently cooled down by a failure of BUU simulations to
describe the energy and angular distribution of neutrons and protons in
low and mid energy domain \cite{T95,Ba95,S96}. Indeed, the
Boltzmann equation is not the full story. As noticed in numerical studies of hard sphere cascade by Halbert \cite{H81} and more general by Malfliet
\cite{M83}, it is disturbing that all dynamical models
rely more or less on the use of the space- and time-local approximation
of binary collisions inherited from the Boltzmann equation. This
approximation neglects a contribution of the collision flux to the
compressibility and the share viscosity which control the hydrodynamic
motion during the reaction. To include the collision flux and other
virial corrections, the non-local character of binary collisions has to
be accounted for. Malfliet also demonstrated that non-local collisions
can be easily incorporated into simulation BUU codes.

In absence of a first principle theory, Malfliet in his pioneering
study, and more recently Kortemeyer, Daffin and Bauer \cite{KDB96},
had to use classical hard-sphere-like non-local collisions which
naturally do not result in a full quantitative agreement with
experimental data. This ad hoc approximation reflects a gap in former
quasi-classical theories of quantum transport: authors either cared
about non-local collisions leaving aside quasiparticle features or vice
versa. Moreover, quantum theories of binary collisions treated non-local
collisions via gradient contributions to the scattering integral
\cite{NTL91,H90} which are numerically inconvenient and thus have never
been employed in demanding simulations of heavy ion reactions. Recent
theoretical studies have filled this gap. Danielewicz and Pratt
\cite{DP96} pointed out that the collision delay can be used as a
convenient tool to describe the virial corrections to the equation of
state for the gas of quasiparticles. Although their discussion is
limited to the equilibrium, it marks a way how to introduce virial
corrections also to dynamical processes. This task was approached in
\cite{SLM96}, where a quasi-classical kinetic equation was derived by a
systematic quasi-classical approximation of non-equilibrium Green's
function in the Galitskii--Feynman approximation. The derived kinetic
equation is suitable for simulations having the quasiparticle
renormalizations in the standard form of Landau's theory and non-local
collisions reminding classical hard spheres.

In this letter we follow the line initiated by Malfliet, however, with
more advanced theoretical and numerical tools. We start from the kinetic
equation from \cite{SLM96} and evaluate non-local collisions from the
Paris potential solving the two-particle T-matrix. 
These non-local corrections are then incorporated into
the QMD simulation. Results show that the non-local corrections bring
the energy distribution of protons closer to experimental values. On a
complementary BUU simulation we enlighten the microscopic mechanism
leading to this improvement.

The scattering integral of the non-local kinetic equation derived in
\cite{SLM96} corresponds to a following picture of a collision. Assume
that two particles, $a$ and $b$, of initial momenta $k$ and $p$ start to
collide at time instant $t$ being at coordinates $r_a$ and $r_b$. Due to
a finite range of the interaction, at the beginning of collision
particles are displaced by $r_b-r_a=\Delta^{\rm be}$. The collision has
a finite duration $\Delta_t$, i.e., it ends at $t+\Delta_t$. During the
collision, both particles move so that their end coordinates differ from
those at the beginning, $r_a'-r_a=\Delta_a$ and $r_b'-r_b=\Delta_b$. The
particle $a$ transfers a momentum $q$ to the particle $b$, therefore
their relative momentum changes from $\kappa=\frac 1 2 (k-p)$ to
$\kappa'=\frac 1 2 (k-p)-q$. Their sum momentum is modified by an
external field acting on the colliding particles during the collision
going from $K=k+p$ to $K'=k+p+\Delta_K$. The same field changes the sum
energy of colliding particles from $E=\epsilon_a+\epsilon_b$ to $E'=
\epsilon_a'+\epsilon_b'=\epsilon_a+\epsilon_b+\Delta_E$.

\begin{figure}
  \psfig{figure=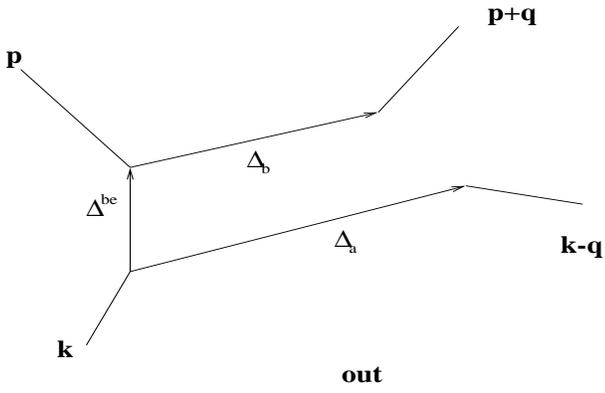,width=8cm,height=5cm}
\caption{A nonlocal binary collision according to Eq. (\protect\ref{delta}).
\label{soft}}
\end{figure}

The values of $\Delta$'s are given by derivatives of the total scattering
phase shift \mbox{$\phi={\rm Im\ ln}T_R(\Omega,k,p,q,t,r)$} of the two-particle T-matrix $T_R$,
\begin{equation}
\begin{array}{lclrcl}\Delta_t&=&{\displaystyle
\left.{\partial\phi\over\partial\Omega}
\right|_{E'}}&\ \ \Delta^{\rm be}&=&
{\displaystyle\left({\partial\phi\over\partial p}-
{\partial\phi\over\partial q}-{\partial\phi\over\partial k}
\right)_{E'}}\\ &&&&&\\ \Delta_E&=&
{\displaystyle\left.-{\partial\phi\over\partial t}
\right|_{E'}}&\Delta_a&=&
{\displaystyle\left.-{\partial\phi\over\partial k}
\right|_{E'}}\\ &&&&&\\ \Delta_K&=&
{\displaystyle\left.{\partial\phi\over\partial r}
\right|_{E'}}&\Delta_b&=&
{\displaystyle\left.-{\partial\phi\over\partial p}
\right|_{E'}}
\end{array}.
\label{delta}
\end{equation}
Note that energy $\Omega$ enters as an independent quantity so that one
needs to know the scattering phase shift out of the energy shell. The
on-shell energy, $\Omega=E'$, is substituted after derivatives are
taken. Figure~\ref{soft} illustrates the nonlocal concept derived in \cite{SLM96}. We like to point out that this concept leads to a continuous trajectory in the kinetic picture replacing real potential scattering. Consequently the energy, momentum, density and angular momentum conservation are conserved including second order quantum virial corrections \cite{SLM96}.

It is our intention to incorporate these features of collisions into the
QMD and BUU simulation codes. The selfconsistent evaluation of all
$\Delta$'s for all collisions would be too demanding. We employ two
kinds of additional approximations. First, following approximations used
within the BUU equation, we neglect the medium effect on binary
collision, i.e., use the well known free-space T-matrix. Second, we
rearrange the scattering integral into an instant but non-local form.
This instant form parallels hard-sphere-like collisions what allow us
to employ computational methods developed within the theory of gases
\cite{AGA95} similarly as it has been done in \cite{KDB96}.

\begin{figure}
  \psfig{figure=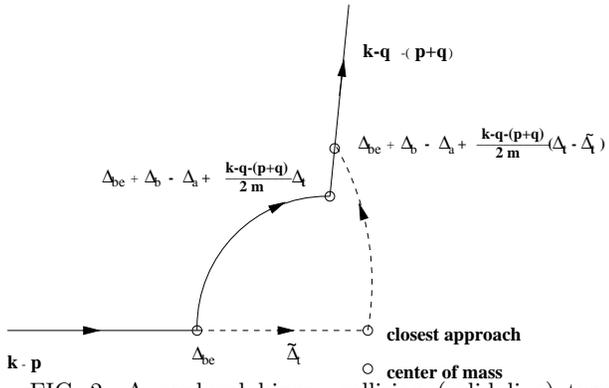,width=8cm,height=5cm}
\caption{A nonlocal binary collision (solid line) together with the szenario of sudden jump at the closest approach.
\label{softc}}
\end{figure}

In the instant approximation we let particles to make a sudden jump at
time $t$ from $r_a$ and $r_b$ to effective final coordinates $\tilde
r_a$ and $\tilde r_b$. These effective coordinates and momenta
$\tilde\kappa$ and $\tilde K$ are selected so that at time $t+\Delta_t$
particles arrive at the correct coordinates, $r_a'$ and $r_b'$, with the
correct momenta, $\kappa'$ and $K'$. Accordingly, in the asymptotic
region, after $t+\Delta_t$, there is no distinction between the
non-instant and instant pictures, which is shown as solid line in figure \ref{softc}. 
This asymptotic condition is naturally
met if one extrapolates the out-going trajectories from known
coordinates and momenta at $t+\Delta_t$ back to the time $t$. Doing so
one finds that the effective coordinates read
\begin{eqnarray}
\tilde r_a&=&r_a'-{k-q\over m}\Delta_t=
r_a+\Delta_a-{k-q\over m}\Delta_t,
\label{da}\\
\tilde r_b&=&r_b'-{p+q\over m}\Delta_t=
r_b+\Delta_b-{p+q\over m}\Delta_t.
\label{db}
\end{eqnarray}
The change of the relative momentum is insensitive to the instant
approximation, $\tilde\kappa=\kappa'=\kappa-q$. The sum momentum and
energy, however, get simplified because during the instant process the
mean-field has no time to pass any momentum and energy to the colliding
pair. Accordingly, $\tilde K=K$ and $\tilde E=E$. In other words, the
momentum and energy gains are naturally covered by the effect of the mean
field on particles during the time interval $(t,t+\Delta_t)$ which, in
the instant picture, is already covered by a free motion. Similarly, in
agreement with the continuity of the center of mass motion, one finds
that $\tilde r_a+\tilde r_b=r_a+r_b$.

When incorporating the displacements into the QMD simulation code, we
have to face the fact that two particles are selected for a collision if they meet at the point of closest approach.
This distance is different from the distance $\Delta^{\rm be}$ required from the equivalent scattering scenario presented in figure \ref{softc} as solid line.We consider now the time required to travel from $\Delta_{\rm be}$ to the distance of closest approach $\tilde\Delta_t={m\over 2 \kappa^2} \kappa \Delta_{\rm be}$ in analogy to \cite{T81}. Within this scenario we are allowed to jump at the point of closest approach to the final asymptotics (\ref{da}) and (\ref{db})whith the additional distance the particle travel during $\tilde\Delta_t$. The effective final coordinates thus have to be evaluated as
\begin{equation}
\tilde r_{a,b}={R_a+R_b \over 2}\mp\Delta^{\rm f},
\label{df}
\end{equation}
with the effective displacement
\begin{equation}
\Delta^{\rm f}=\frac 1 2 \Delta^{\rm be}-\Delta_a+{k-q\over m}(\Delta_t-\tilde\Delta_t).
\label{dd}
\end{equation}
Since the center of mass does not jump in the collision, the final
displacement can be also written in an alternative way, $\Delta^{\rm f}
=\frac 1 2 \Delta^{\rm be}+\Delta_b-{p+q\over m}(\Delta_t-\tilde\Delta_t)$. The
non-local corrections are thus performed as follows. When the collision
is selected, we evaluate $\Delta^{\rm f}$ from (\ref{dd}) and
(\ref{delta}), redisplay particles into $\tilde r_a$ and $\tilde r_b$
and continue with the simulation.

At this point it is possible to establish a connection of the present
theory to the hard-sphere-like corrections used by Malfliet \cite{M83}
and Kortemeyer, Daffin and Bauer \cite{KDB96}. For hard spheres of the
diameter $d$, the phase shift has a classical limit $\phi=\pi-|q|d$
which gives $\Delta_{a,b}=0$ and $\Delta^{\rm f}=\Delta^{\rm be}=
{q\over|q|}d$. The displacement thus has the same amplitude $d$ for all
binary collisions and points in the direction of the transferred
momentum, as it is known from the Enskog equation \cite{CC90}. In the
present theory, an amplitude of displacement (\ref{dd}) depends on the
selected channel, $\Delta^{\rm f}(k,p,q)$, and the direction does not
coincide with the transferred momentum,
\begin{equation}
\Delta^{\rm f}={q\over|\kappa|}H_2+{k-p-q\over|\kappa|}H_1.
\label{cc}
\end{equation}
The second term is perpendicular to the transferred momentum and stays
in the collision plane. The notation of coefficients $H_{1,2}$ is
identical to the corresponding interface subroutines which can be
obtained from authors.

\begin{figure}
\parbox[t]{7cm}{\psfig{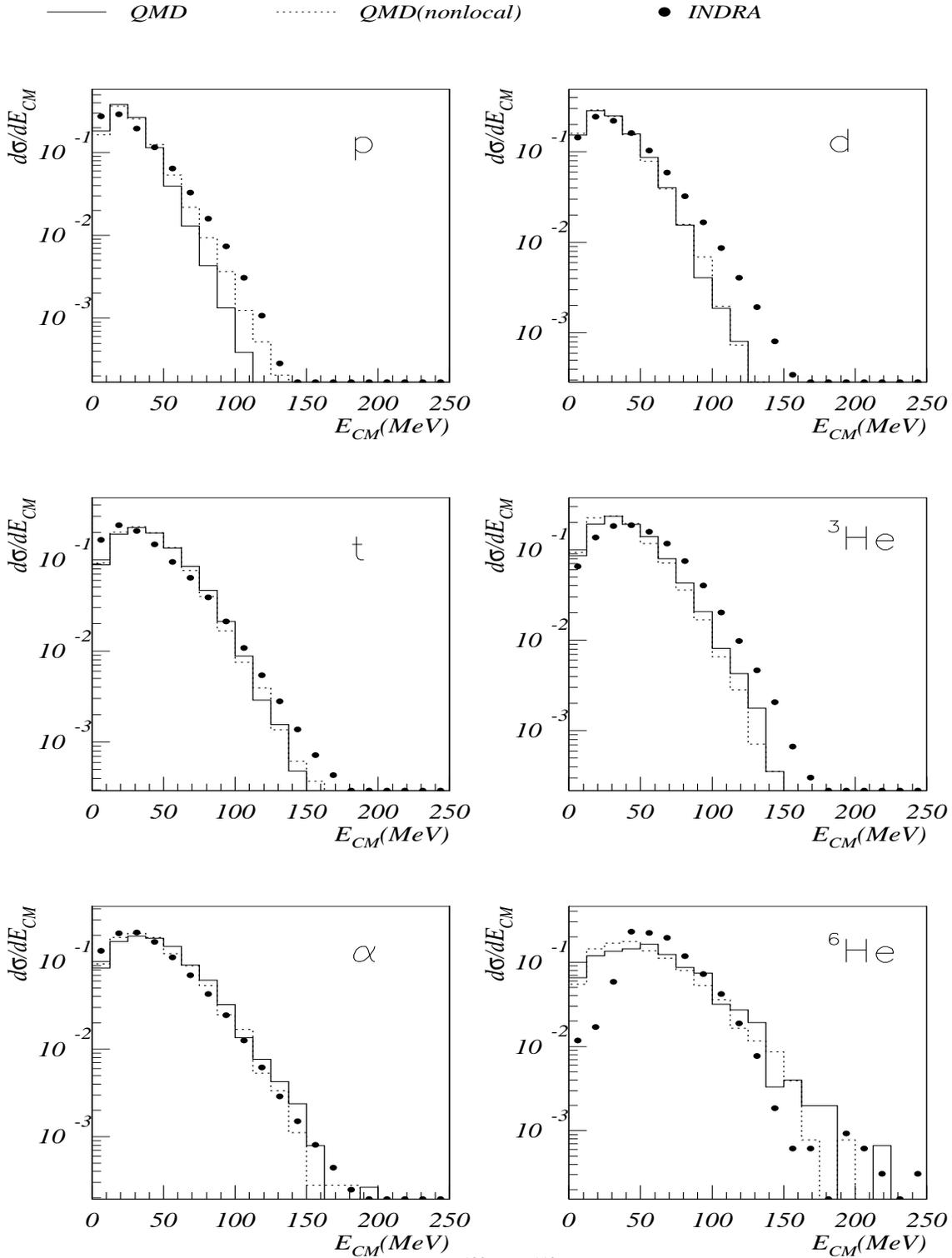}}
\caption{The particle spectra for central collision of
$^{129}$Xe$\rightarrow$$^{119}$Sn at $50$~MeV/A with and without
non-local corrections. The data are extracted from recent INDRA
experiments \protect\cite{INDRA}. The non-local corrections bring the spectrum
of the protons towards the experimental values leaving the
clusters almost unchanged.
\label{spek3}}
\end{figure}

In order to investigate the effect of non-local shifts on realistic
simulations of a heavy ion reaction, we have evaluated $\Delta^{\rm f}$
from the two-particle scattering T-matrix $T^{R}$ in the Bethe-Goldstone 
approximation \cite{D84,SLM96} using the separable Paris potential \cite{HP84}.
The comparison of the shifts calculated for different potentials concerning partial wave coupling up to D-waves can be found in \cite{MLSK98}. 
We have incorporated these shifts into
a QMD code for the central collision of
$^{129}$Xe$\rightarrow$$^{119}$Sn at $50$~MeV/A. Figure~\ref{spek3}a
shows the exclusive proton spectra subtracting the protons bound in
clusters. This procedure is performed within a spanning tree model which
is known to describe a production of light charged cluster in a
reasonable agreement with the experimental data, Figs.~\ref{spek3}b-f.
Within the local approximation, however, the remaining distribution of
high-energy protons is too low to meet the experimental values. As one
can see, the inclusion of non-local collisions corrects this shortage of
the QMD simulation. As demonstrated in Fig.~\ref{spek3}, productions of
light clusters are rather insensitive to the non-local corrections. This
also shows that the improvement of the proton production is not on cost
of worse results in other spectra.

\begin{figure}
\parbox[t]{6cm}{
\psfig{file=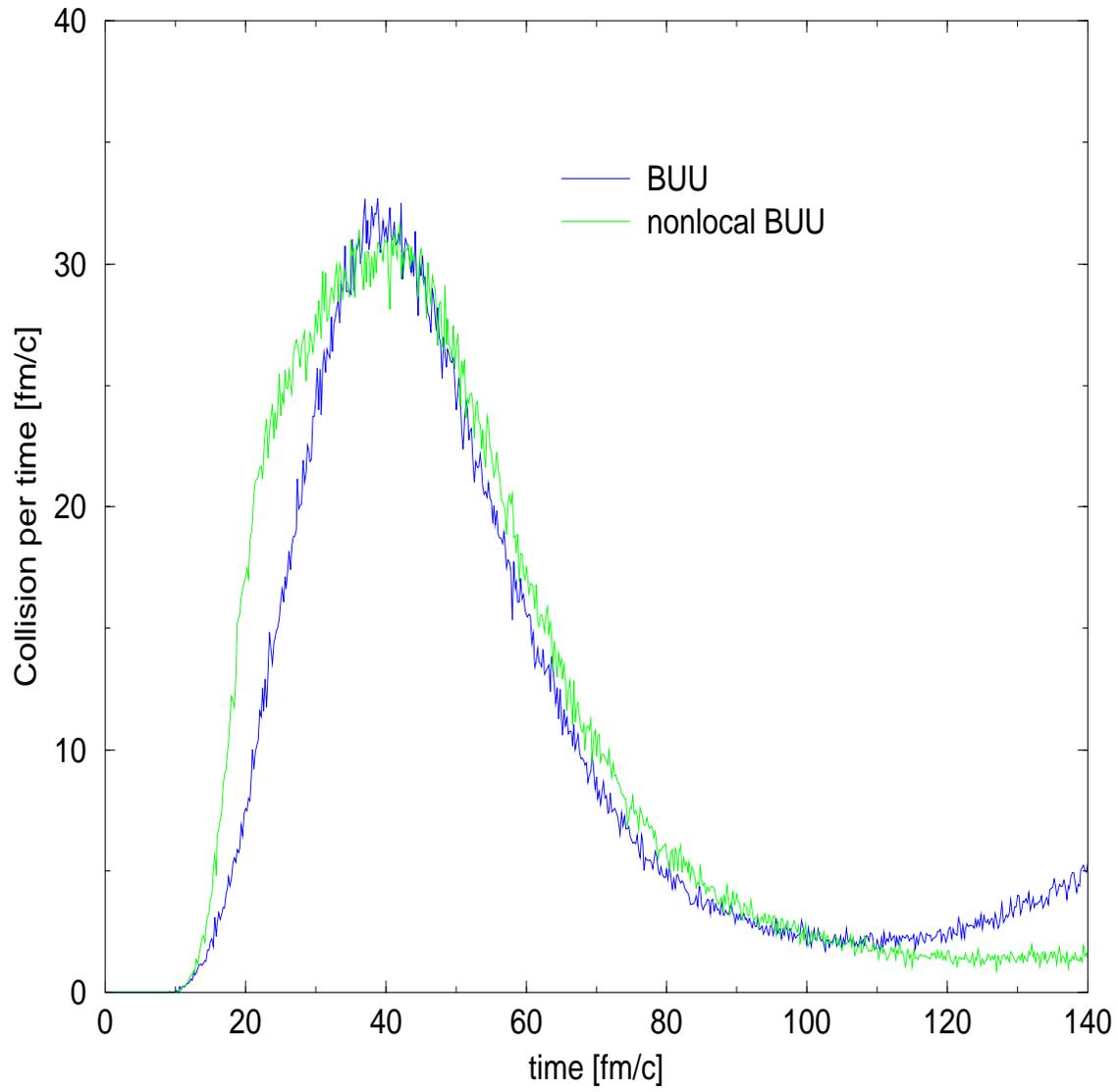,width=15cm,height=15cm,angle=-90}}
\caption{The number of collisions per time with and without non-local
collisions within a BUU simulation of the same reaction as in figure
(\protect\ref{spek3}).
\label{n}}
\end{figure}

A microscopic mechanism leading to the increase in the high-energy part
of the particle spectrum can be traced down to an enhancement of the
number of collisions at the pre-equilibrium stage of the heavy ion
reaction demonstrated in Fig.~\ref{n} for the BUU simulation of the same
reaction. This enhancement gives rise to an immediate proton production
which itself translates into a high energetic spectra. In other words,
the strong production of the high-energy protons follows from the
pre-equilibrium emission of particles. The BUU simulation also shows
that non-local corrections are important namely in the early stage of
reaction well before most of light clusters are formed. It explains why
the production of protons is affected while the formation of light
clusters is nearly untouched by the non-local corrections.

In summary, as documented by the improvement of the high-energy proton
production, the non-local treatment of the binary collisions brings a
desirable contribution to the dynamics of heavy ion reactions. According
to an experience from the theory of gases, one can also expect a vital
role of non-localities in the search for the equation of state of the
nuclear matter. It is encouraging that the non-local corrections are
easily incorporated into the BUU and QMD simulation codes and do not
increase computational time.

\medskip\noindent
We thank the INDRA collaboration for the use of data prior to
publication. J.Aichelin and W. Bauer are thanked for the well documented
QMD and BUU codes. This work was supported from Czech Republic, the GACR
Nos.~202960098 and 202960021, and GAAS Nr. A1010806, and Germany, the
BMBF, Nr. 06R0884, and the Max-Planck-Society.


\begin{thebibliography}{10}

\bibitem{BD88}
G.~F. Bertsch and S.~D. Gupta, Phys. Rep. {\bf 160},  189  (1988).

\bibitem{SG86}
H. St{\"o}cker and W. Greiner, Phys. Rep. {\bf 137},  277  (1986).

\bibitem{A91}
J. Aichelin, Phys. Rep. {\bf 202},  235  (1991).

\bibitem{T95}
J. T{\~o}ke and et. al., Phys. Rev. Lett. {\bf 75},  2920  (1995).

\bibitem{Ba95}
S. Baldwin and et. al., Phys. Rev. Lett. {\bf 74},  1299  (1995).

\bibitem{S96}
W. Skulski and et. al., Phys. Rev. C {\bf 53},  R2594  (1996).

\bibitem{H81}
E.~C. Halbert, Phys. Rev. C {\bf 23},  295  (1981).

\bibitem{M83}
R. Malfliet, Nucl. Phys. A {\bf 420},  621  (1983).

\bibitem{KDB96}
G. Kortemeyer, F. Daffin, and W. Bauer, Phys. Lett. B {\bf 374},  25  (1996).

\bibitem{NTL91}
P.~J. Nacher, G. Tastevin, and F. Laloe, Ann. Phys. (Leipzig) {\bf 48},  149
  (1991).

\bibitem{H90}
M. de~Haan, Physica A {\bf 164},  373  (1990).

\bibitem{DP96}
P. Danielewicz and S. Pratt, Phys. Rev. C {\bf 53},  249  (1996).

\bibitem{SLM96}
V. {\v S}pi{\v c}ka, P. Lipavsk{\'y}, and K. Morawetz, Phys. Lett. A {\bf 240},
   160  (1998).

\bibitem{AGA95}
F.~J. Alexander, A.~L. Garcia, and B.~J. Alder, Phys. Rev. Lett. {\bf 74},
  5212  (1995).

\bibitem{T81}
W. Thirring,  in {\em Classical Scattering Theory}, Vol.~XXIII of {\em Acta
  Physica Austriaca, Suppl.}, edited by H. Mitter and L. Pittner
  (Springer-Verlag, Wien, 1981), p.\ 3.

\bibitem{CC90}
S. Chapman and T.~G. Cowling, {\em The Mathematical Theory of Non-uniform
  Gases} (Cambrigde University Press, Cambridge, 1990), third edition Chap. 16.

\bibitem{D84}
P. Danielewicz, Ann. Phys. (NY) {\bf 152},  239  (1984).

\bibitem{HP84}
J. Heidenberger and W. Plessas, Phys. Rev. C {\bf 30},  1822  (1984).

\bibitem{MLSK98}
K. Morawetz, P. Lipavsk{\'y}, V. {\v S}pi{\v c}ka, and N. Kwong, Phys. Rev. C
  (1998), sub.

\bibitem{INDRA}
Private communication from INDRA collaboration.

\end{thebibliography}

\onecolumn

\end{document}